\documentclass[9pt,twocolumn,twoside]{osajnl}
\usepackage{graphicx}
\usepackage{subfigure}
\usepackage{subfiles}
\usepackage{caption}
\usepackage{amsmath,amssymb,epsfig,hyperref}
\usepackage{enumerate}
\usepackage{float}
\usepackage{gensymb}
\newcommand{\figref}[1]{Fig.~\ref{#1}}

\journal{ao} % Choose journal (ao, aop, josaa, josab, ol)

\title{Spatially correlated photonic qutrit pairs using pump beam modulation technique}

\author{Debadrita Ghosh}
\author[2,3]{Thomas Jennewein}
\author[4]{Piotr Kolenderski}
\author[1,*]{Urbasi Sinha}

\affil[1]{Raman Research Institute,Sadashivanagar,Bengaluru-560080}
\affil[2]{Institute for Quantum Computing,University of Waterloo,200 University Avenue West,Waterloo,Ontario N2L3G1,Canada}
\affil[3]{Department of Physics and Astronomy,University of Waterloo,Waterloo, Ontario,Canada}
\affil[4]{Faculty of Physics, Astronomy and Informatics, Nicolaus Copernicus University, Grudziadzka 5, 87-100 Toru\'{n}, Poland}

\affil[*]{Corresponding author: usinha@rri.res.in}

\begin{abstract}
Higher dimensional quantum systems have a very important role to play in quantum information, computation as well as communication. While the polarization degree of freedom of the photon is a common choice for many studies, it is restricted to only two orthogonal states, hence qubits for manipulation. In this paper, we theoretically model as well as experimentally verify a novel scheme of approximating photonic qutrits  by modulating the pump beam in a spontaneous parametric down conversion process using a three-slit aperture. The emerging bi-photon fields behave like qutrits and are found to be highly correlated in the spatial degree of freedom and effectively represent spatially correlated qutrits with a Pearson coefficient as high as 0.9. In principle, this system provides us a scalable architecture for generating and experimenting with higher dimensional correlated qudits.
\end{abstract}

\setboolean{displaycopyright}{false}

\begin{document}

\maketitle

\section{Introduction}
The ability to generate and detect correlated photon pairs by means of spontaneous parametric down conversion (SPDC) offers lots of experimental possibilities ranging from fundamental tests of quantum theory \cite{Sinha2010} to practical applications in quantum information \cite{Nielsen2000} and quantum communication \cite{Pugh2016}. The influence of the pump beam spatial mode on the characteristics of the resulting photon pairs has been analyzed in \cite{Monken} and in terms of optimization of the collecting of the photon pairs in single mode fibres in \cite{Gajewski2016}. Recently it was analysed how the pump beam spatial mode influences the spatial characteristics of the resulting photon pairs \cite{Pugh2016}. In Ref.~\cite{Vicuna-Hernandez2016} the authors analyze the case where the pump beam is prepared in Bessel-Gauss mode. In turn, Pugh \textit{et al}  \cite{Pugh2016} show the way to control the pump beam for the application for long distance quantum communication with satellites. 

Along these lines, there is an extensive research towards encoding of quantum information in a single photon's degree of freedom. Spatial degree of freedom offers a variety of possibilities such as orbital angular momentum \cite{Mair2001,Molina-Terriza2007,Calvo2007,Malik2012} or spatial qudits \cite{Neves2004,Neves2005,Neves2007,Taguchi2009}. The latter framework proved to be useful when testing fundamentals of quantum theory \cite{ Sinha2015, Kolenderski2011, Sawant2014}.  In earlier work \cite{Kolenderski2011}, it has been shown that a triple slit aperture placed in the path of a photon leads to the generation of a spatial qutrit. Note that in practice the transmission of the system, which is fundamentally very low, limits the robustness in terms of counting rates.
The use of calcite beam displacer can reduce this problem for a qubit implementation \cite{Kolenderski2014,Kolenderski2013a}. However the scaling towards higher dimensional states is challenging within this framework.

Here we study the case where the pump beam is prepared such that its spatial mode resembles three slits, and the spatial structure is carried through to the resultant signal and idler photons. We investigate the correlations between the signal and idler photons in the spatial domain and effectively verify our spatially correlated qutrits have a Pearson coefficient as high as $0.9$.

In this work, we show that the structure of the pump field is preserved in the process of SPDC and we find very convincing match between theoretical predictions and experimental results for the correlation between the signal and the idler photons. We also lay down the recipe for the optimal choice of experimental parameters which can be exploited in future architecture to maximize the correlations. 

%\end{document}

\section{Three-slit-based qutrits}
\begin{figure}[h]
\centering
	\includegraphics[width=\columnwidth]{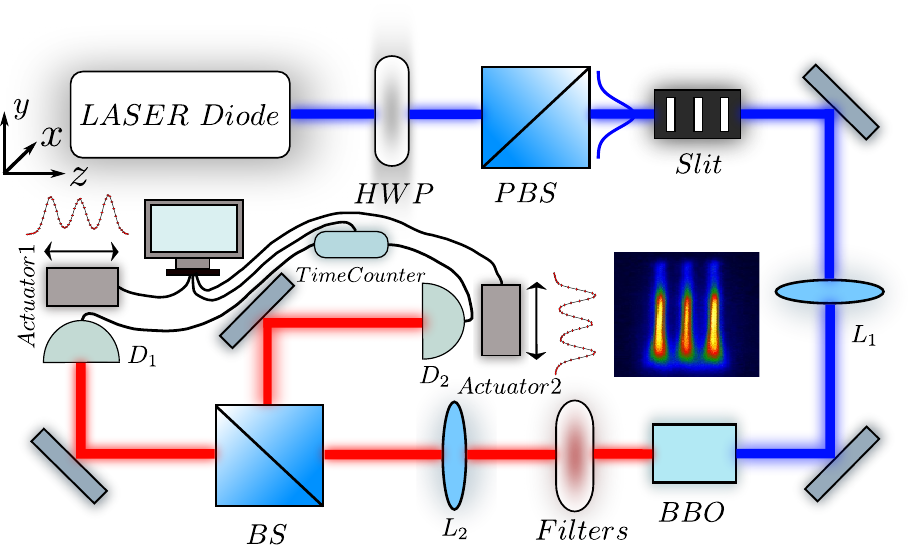}
	\caption{Schematic of the experimental set-up. Horizontal pump beam is made incident on a triple slit aperture. Lens $L_1$ is used to transfer the image of the pump beam on the Type-1 BBO crystal. After appropriate filtering of the blue pump beam, another lens $L_2$ is used to transfer the signal and idler spatial profiles to actuated detectors placed on either side of a beam splitter. The spatial profiles of the signal and idler photons are measured using detectors $D_1$ and $D_2$ and the spatial correlation is measured using an appropriate co-incidence logic unit.}
	\label{fig:setup}
\end{figure}

Let us consider a setup depicted in \figref{fig:setup} consisting of a Type-1 non-linear crystal. We are using the collinear geometry for our down conversion process. The pump beam spatial mode is prepared by transferring a Gaussian beam through a set of three slits (centered at the middle slit) and imaging the result on the crystal using a lens in the 2f - 2f configuration. The coordinate system orientation is chosen such that the propagation is along $z$ axis, and $x$ and $y$ axes are parallel to the shorter and longer length of the slits, respectively. 

We model the pump beam as three box functions weighted by a Gaussian as the pump profile in the crystal to generate the bi-photon qutrits and their resultant correlations so that the simulations ended in finite time. 

We followed \cite{Boeuf2000} to solve for phase matching in the Type I SPDC process. The crystal that we have used is BBO with a cut angle of $29.3 \degree$ for non collinear SPDC process. 
For type I SPDC, we define the phase mismatch as 

\begin{align}
\Delta \vec{k} = \vec{k_p}(\omega_p, \alpha, n_e(\theta)) - \vec{k_s}(\omega_s, \alpha, n_o) - \vec{k_i}(\omega_i, \alpha, n_o).
\end{align} 
Here, $\omega_p, \omega_s$ and $\omega_i$ refer to the frequency of the pump, signal and idler photons respectively, and $n_e$ and $n_o$ refer to extraordinary and ordinary refractive indices. The pump wave vector inside the crystal depends on the angle $\alpha$ it makes with the optic axis of the crystal. 

The SPDC process considered here is $e \rightarrow oo$. 
For collinear degenerate SPDC, we expect  $\Delta \vec{k} =0$. Using parameters of refractive index from BBO Sellmeier equations, we have the collinear phase matching at  $28.81\degree$.

Given a phase mismatch $\Delta \vec{k}$, we assign the intensity weight to it as 
\begin{align}
|sinc(\Delta k_x L_x) sinc(\Delta k_z L_z)|^2,
\end{align}
where  $L_x$ is the transverse length of the crystal and $L_z$ is the thickness of the crystal along beam propagation direction. We have ignored the transverse length along the height i.e. y-axis for ease of computation.

Following \cite{Boeuf2000, Kolenderski2009}, we can find the Hamiltonian for the SPDC process where the pump is treated classically and the signal and idler are treated as perturbations. For degenerate down-conversion $\omega_p = 2 \omega_s = 2 \omega_i$, and the probability amplitude to get a coincidence for degenerate photons generated from collinear SPDC will be:
\begin{align}
\mathcal{A}  \propto \int_{V} d^3 r \int d k_s \int d k_i  A_p(x,y,z) \exp(i (\vec{k_p}-\vec{k_s}-\vec{k_i}).\vec{r})
\end{align}
Here, $A_p(x,y,z)$ is the amplitude of the pump as a function of space. $z$ is assumed to be the direction of propagation of pump, $\vec{k_p},\vec{ k_s}$ and $\vec{k_i}$ are the pump, signal and idler wave vectors respectively. The integral over $dk_s$ depends on the angle that the detector subtends to the crystal. For point detectors we have a single value of $\vec{k_{s,i}}$. However, the limits of integration for finite sized detectors for the $\vec{k_s}$ will be of the form $\vec{k^\mu_s}\pm\vec{ k^\Delta_s}$. Here $\vec{k^\mu_s}$ is determined by the position of the detectors and $\vec{k^\Delta_s}$ is determined by the size of the detectors.

We now attempt to compute $\mathcal{A}$ numerically. However, integration of the complex exponential is numerically cumbersome and does not converge. Hence we assume that $\vec{k_p}$ does not depend on position inside the crystal, so that we can integrate the exponential independently and replace it with a sinc function.\\
We use the approximation
\begin{align}
\mathcal{A}(k_s, k_i) = \int_{V} d^3 r  A_p(x,y,z) \Pi_{i=x,y,z}sinc(\Delta k_i(\Delta r) L_i)
\label{corrProper}
\end{align}
Here $\Pi_{i=x,y,z}sinc(\Delta k_i(r) L_i)$ should be interpreted as the weight associated with a point classical pump and here the phase matching is weighted by the length of the crystal. Then we consider that the pump is composed of several such point pumps and we assume that different points in the crystal are independent as far as wave vectors are considered. Note that phase mismatch depends on $\Delta r$, the difference between crystal and detector positions (and not just on the crystal position).

In the simulation code, we integrate over two coordinates (x and z). We are interested in the transverse correlation (along x). The height of the slit is assumed to be infinity, pump wave vector is assumed to be along z-direction only, assuming a thin crystal and each point in the crystal is assumed to  have SPDC independently with the weight $sinc(\Delta k_i (\Delta r) L_i)$.

We follow the following steps to compute $|\mathcal{A}(k_s, k_i)|^2$.
\begin{itemize}
\item Compute $\Delta k$ for a pair of signal and idler points and for a given point on the crystal.
\item Compute the point to point correlation as the product of intensity of the pump at the given crystal point $|A_p|^2$ and the weight function $|\Pi_{i=x,y,z}sinc(\Delta k_i(r) L_i)|^2$
\item Integrate over the signal/idler position about the mean detector position and with range equal to the size of detector.
\item Repeat the above for different idler points to get the spatial correlation between a signal position and the entire idler profile.
\item Sum the idler profile computed for different signal positions to get the signal profile. We moved the signal detector with a step size of $3 \mu m$ to attain convergence in this code.
\end{itemize}

The above simulations were carried out for different choice of simulation parameters to obtain an optimal set of parameters that gives us a high value of Pearson correlation coefficient $\rho$ \cite{Pearson1895}. In these simulations, we have assumed point detectors to enable faster simulations. In the result of the numerical simulation we get an approximation of the probability distribution of the coincidence detection of signal photon at position $x_1$ and idler photon at position $x_2$. Based on the numerical data one can easily compute correlation coefficient as a ratio of co-variance over a product of respective variances for signal and idler photon detector positions: $\rho =<x_1 x_2>/\sqrt{<x_1^2><x_2^2>}$.  

We perform simulations for both Type I and Type II SPDC using the same crystal, lens and slit parameters and found that the Pearson coefficient for Type II process is  $0.985(2)$ whereas for Type I, it is $0.966(2)$. We selected an experiment using a Type I crystal as we were concerned that transverse walk off which could happen to one of the two orthogonal polarization exhibited by signal and idler photons could mask the otherwise high correlations that have been predicted from theory.

We have chosen the three slit system to have slit width of $30 \mu m$, inter-slit distance of $100 \mu m$, slit height of $300 \mu m$. For this, we have chosen an incident pump beam with a Gaussian RMS width of $300 \mu m$. This is an optimal choice as we wish to strike a balance between throughput and uniform distribution. For a wider beam width, the number of singles and coincidence counts is expected to drop significantly whereas a smaller beam width would result in the three slits not all being centered close to the peak of the Gaussian.

We have done the simulations for each of the slit parameters keeping other parameters constant and ascertained that the Pearson coefficient is high for our final choice. Table 1 shows the comparison between different slit and crystal parameters in terms of the Pearson coefficient. Further details of these parameter optimization simulations are discussed in the Appendix.

\begin{table}
\begin{center}
 \begin{tabular}{||c| c| c| c||} 
 \hline
 $w$ in $\mu m$ & $d$ in $\mu m$ & $L_z$ in mm & Pearson coefficient\\ [0.5ex] 
 \hline\hline
 30 & 50 & 10 & 0.872393 \\ 
 \hline
 30 & 100 & 10 & 0.965584 \\
 \hline
 30 & 200 & 10 & 0.992088 \\
 \hline
  &  &  & \\
 \hline
 5 & 100 & 10 & 0.964352 \\ 
 \hline
 10 & 100 & 10 & 0.966075\\
 \hline
 30 & 100 & 10 & 0.965584\\
 \hline
 40 & 100 & 10 & 0.966569\\
 \hline
  & & & \\
\hline
  30 & 100 & 5 & 0.975763\\
  \hline
  30 & 100 & 10 & 0.965584\\
  \hline
  30 & 100 & 100 & 0.88873\\[1ex]
\hline
\end{tabular}
\caption{Table comparing the Pearson coefficients varying slit width $w$, inter-slit distance $d$, crystal longitudinal length $L_z$ respectively.}
\end{center}
\end{table}

%\end{document}

\section{Experiment}

The Type I BBO crystal is cut for non-collinear phase matching at $405$ nm to $2 \times 810$ nm at $29.3 \degree$ which translates to collinear phase matching at $28.8 \degree$. The parameter of the slits and pump beam are as mentioned above and the focal length of the lens performing image transfer from slit plane to center of non linear crystal is $146 mm$ for blue incident beam. The focal length of the lens performing the image transfer of the signal and idler photons to the detector plane is $150 mm$ for IR wavelength. 

The detectors D1 (D2) are mounted on motorized stages (actuated with stepper motorized actuators ZST225B and ZST213 from Thorlabs respectively) allowing control of their position in the plane orthogonal to the propagation of signal (idler) photon. The accuracy of the motors is sub micron level. We measured single counts and coincidences of detection at both detectors with a coincidence time window of 1024 ps using FPGA electronics (UQD LOGIC-16). We scanned the characteristic range in the direction orthogonal to the slits' longer dimension with both the detectors. While one of the detectors (D1) scanned the signal photon spatial profile, D2 scanned the idler photon profile. By keeping D1 fixed at different positions of the signal profile, we scanned the detector D2 to yield correlations between the signal position and the entire idler profile. We decided on 13 fixed detector positions for sufficient statistics. These 13 positions correspond to peaks, dips and asymmetrically chosen slope positions to give us maximum information as per a Nyquist sampling criterion. For each D1 fixed position, D2 was moved with a step size of $10 \mu m$. The data acquisition time was $180$ sec at each point, resulting in one complete D2 run taking close to three hours. We repeat the measurement 5 times for better averaging. Thus, for each fixed signal position, we take close to 15 hours to generate the idler profile and resultant correlations. 

\section{Results}
\begin{figure}[H]
%		\subfigure[]{\includegraphics[width=\columnwidth]{grid}}
\centering
	\subfigure[Coincidence map]{\includegraphics[width=0.9\columnwidth]{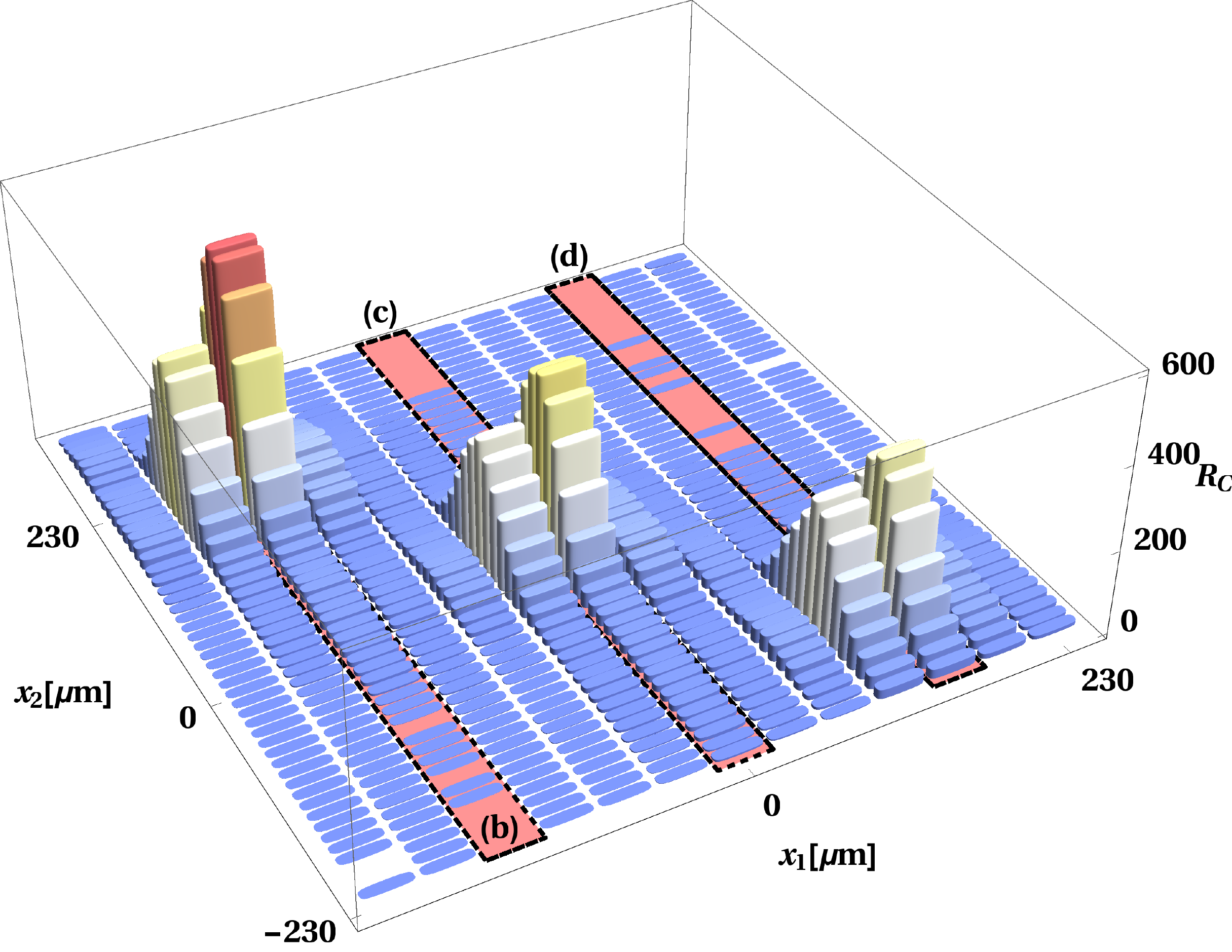}}
	\begin{tabular}{c c}
		\subfigure [D1 at the slit A, D2 moving]{\includegraphics[width=0.45\columnwidth]{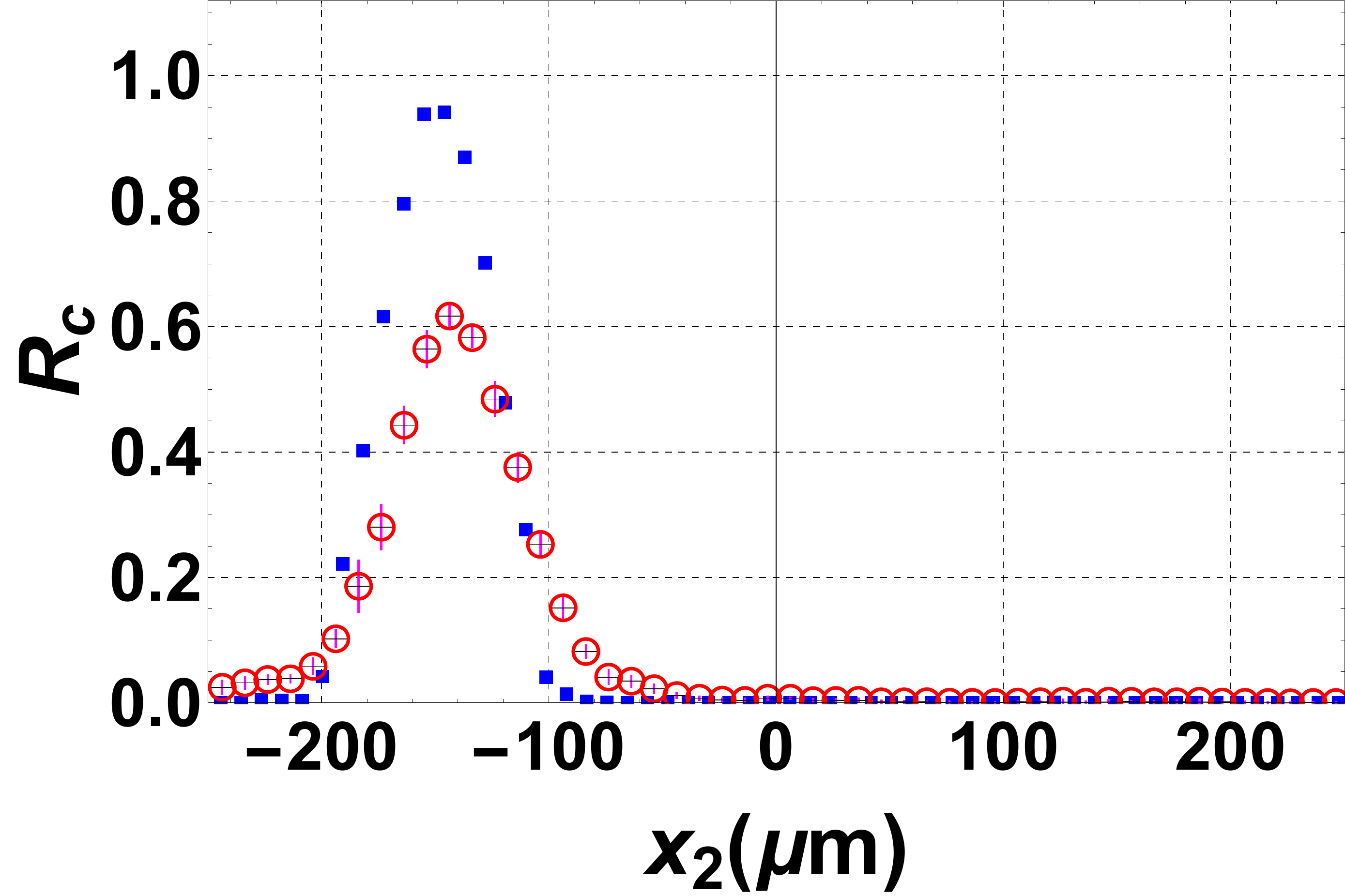}} &
		\subfigure[D1 at the slit B, D2 moving]{\includegraphics[width=0.45\columnwidth]{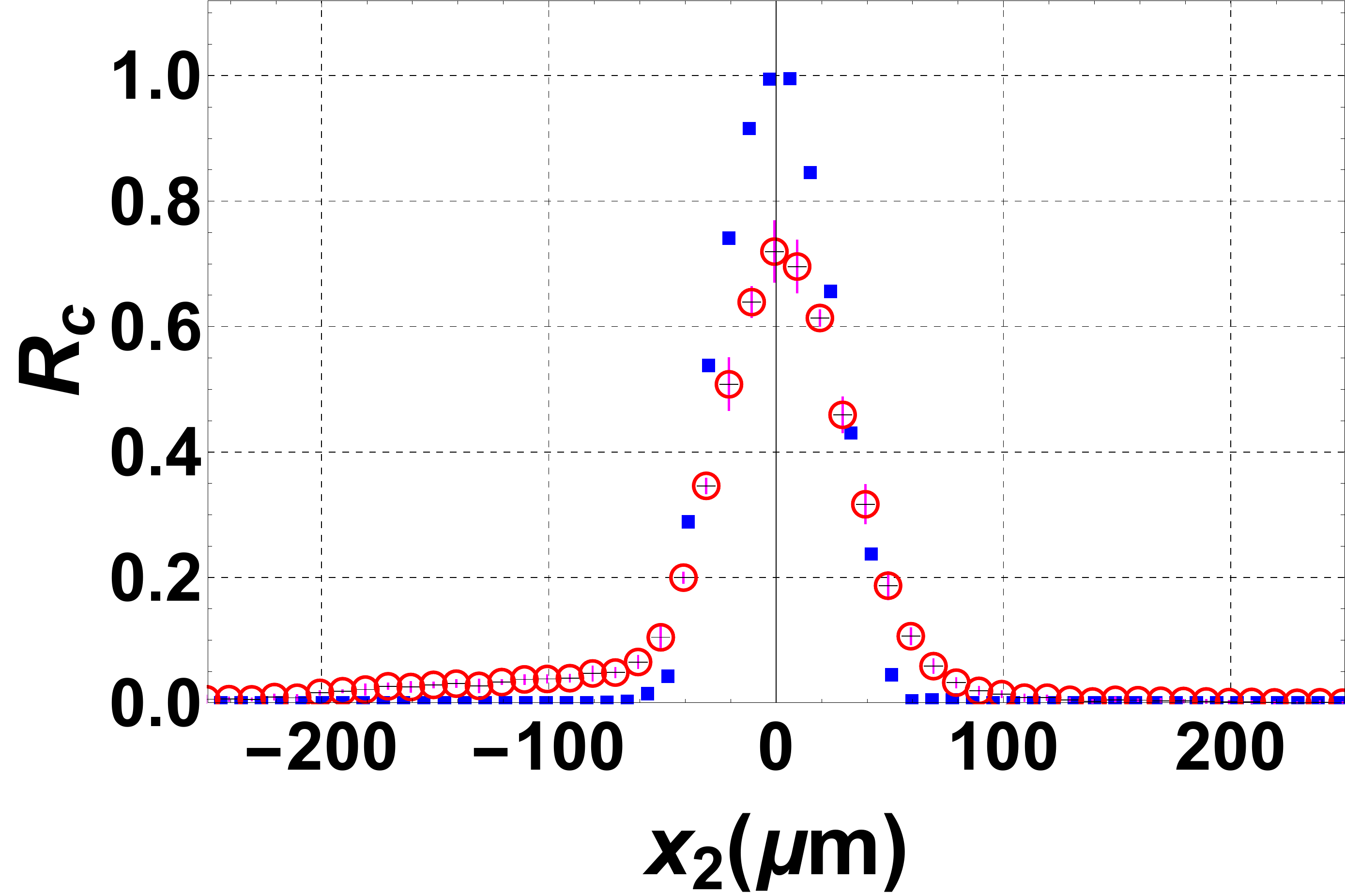}}\\
		\subfigure[D1 at the slit C, D2 moving]{\includegraphics[width=0.45\columnwidth]{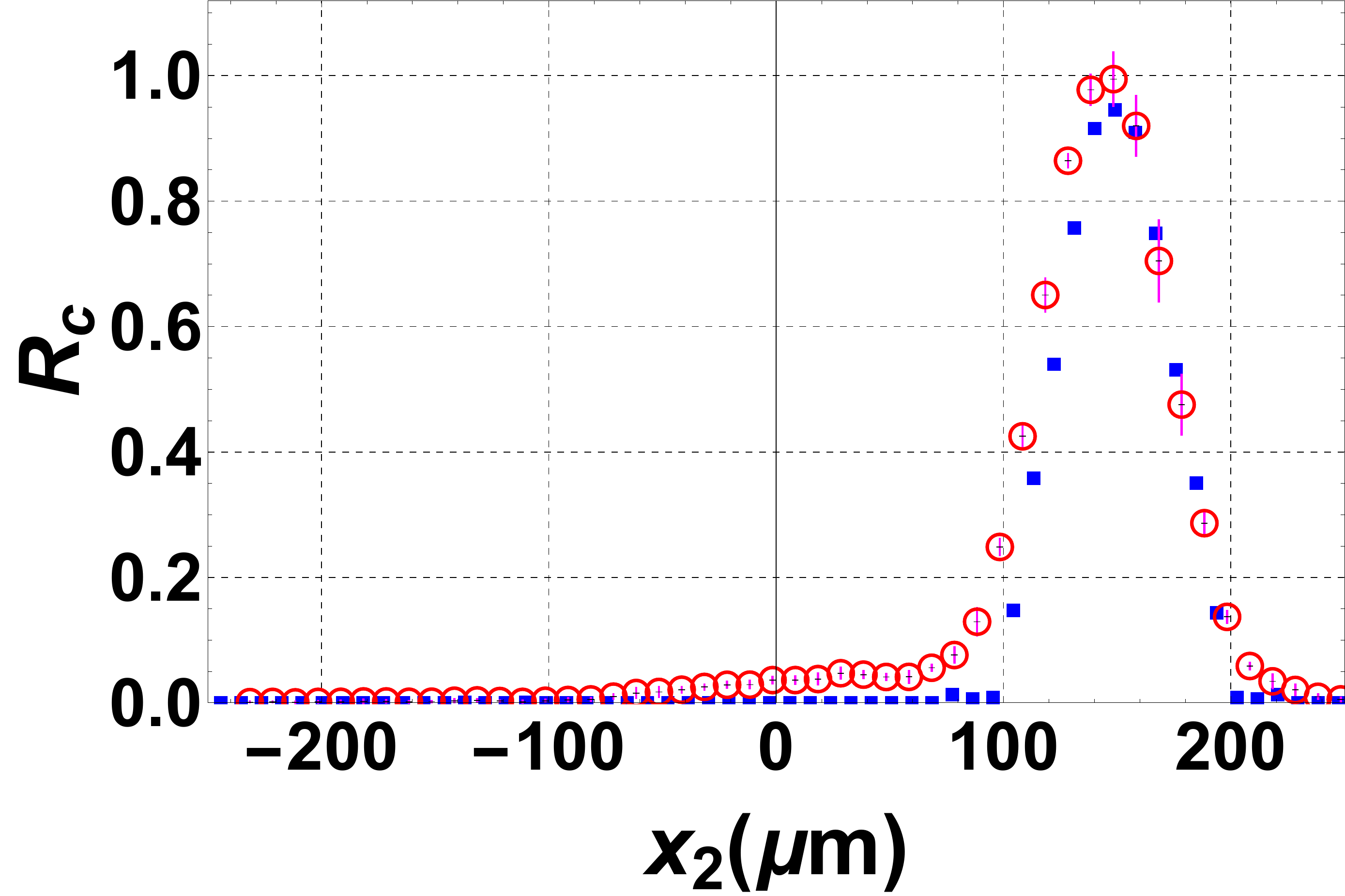}} &
		\subfigure[D1 moving]{\includegraphics[width=0.45\columnwidth]{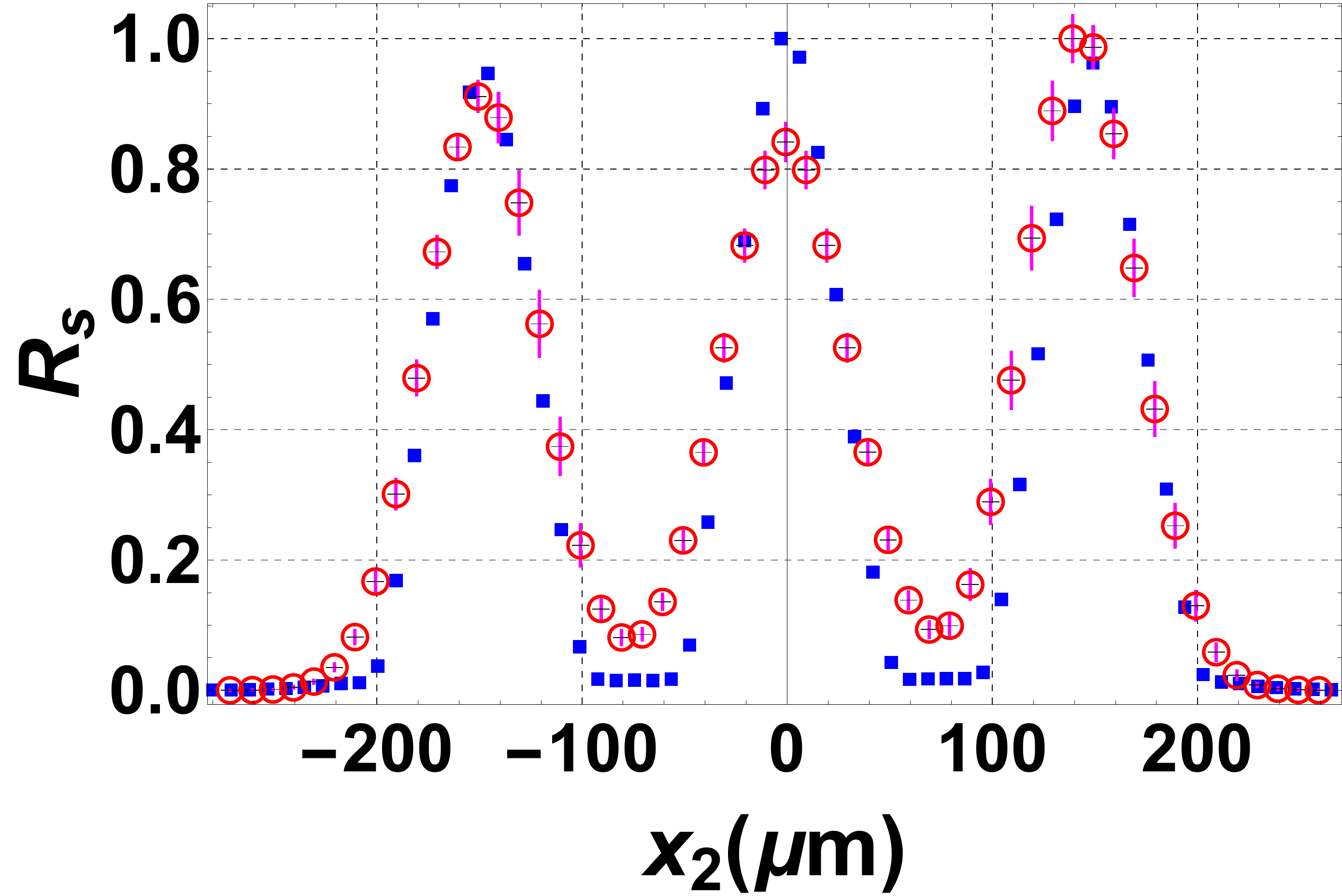}}
	\end{tabular}
	\caption{
	(a) Coincidence counts, $R_C$ measured as a function of position of detectors D1 and D2. A comparison between experimental (Red circle) and theoretically predicted coincidence counts (Blue rectangle) when detector D1 is fixed at peaks of slit (b) A, (c) B, (d) C and the detector D2 is scanning. (e) Single counts, $R_S$ measured at detector D1. Each data point has a measurement time of 3 minutes. }
	%\label{fig:resultsone}
	\label{fig:results}
\end{figure}

The result of coincidence counting is plotted in \figref{fig:results}(a). The experimental correlations between signal and idler positions are appropriately captured by the measurement of coincidences. For the measure of correlation we take the Pearson's coefficient. Here the random variables are detectors' positions, $x_1$ and $x_2$ and the probability distribution is estimated directly by our measurement. The probability of getting a coincidence detection at positions $x_1$ and $x_2$ is proportional to the coincidence counts measured and depicted in \figref{fig:results}(a). The estimate of the Pearson's correlation coefficient is $0.9(2)$. We estimated the uncertainty of the coefficient by simulating $10^5$ probability distributions based on the measured statistics and assuming Poisson statistics of the counts. 

Red dots in \figref{fig:results}(b), (c) and (d) correspond to coincidences, $R_c$, measured as the idler detector is moved while the signal detector is kept fixed at first peak (slit A), second peak (slit B) and third peak (slit C) respectively. Error bars for both position and number uncertainty have been included. The blue lines represent the theoretically simulated correlation profiles.

There are three ways in which one can generate the commensurate theory graphs with respect to the modeling of the pump profile at the crystal. Two methods, discussed in the Appendix , involve using a detailed image transfer formalism or using three box functions weighted by a Gaussian. In order to capture the correlations due to the actual pump beam profile that has been transferred to the crystal, we measured the pump beam profile that is transferred to the position of the center of the crystal using a lens in 2f-2f configuration and used this profile itself to generate the signal and idler profiles as well as calculating the spatial correlations between the two profiles. Slight difference in magnification between the experimental and theoretically simulated profiles has been accounted for in the simulations. An example of a pump beam profile at the crystal position is given in the supplementary material. The use of the experimentally measured pump profile to generate the correlation has the advantage that any non-idealness that may exist in the experiment in terms of alignment or otherwise would then be captured in the theory and the comparison would not suffer from comparing experimental data with ideal theory conditions. 

While \figref{fig:results}(a) shows $R_c$ for all measured detector positions, in \figref{fig:results}(b), (c) and (d), we have chosen to highlight the correlations at the peak positions in a 2-D format to enable representation of the error bars  in terms of position and number uncertainty and also indicate the extent of overlap between theoretical predictions and experimental results.   \figref{fig:results}(e) shows the comparison between the single photon counts $R_s$ measured as a function of detector position for the signal photon and the theoretically generated profile. Experimental and theoretically generated $R_c$ and $R_s$ have been appropriately normalized by their respective maxima.

%\end{document}

Actual measurements indicate that the slit widths are not exactly $30 \mu m$ and $100 \mu m$ respectively as taken in simulations. Slit C is the widest at about $36 \mu m$, slit B the thinnest at around $29 \mu m$ while slit A is in between at around $33 \mu m$. This is reflected in the singles profile where one can see that slit C has maximum counts while B has minimum. The data has been normalized with the maximum (here C) counts. As the coincidence counts are in principle proportional to  the singles, when fixed detector is at position of slit C, we do measure maximum coincidences there. However, in measurement of coincidences, a couple more factors also come in. While the $R_S$ is generated with a single detector (D1) motion controlled by one actuator, $R_C$ involves both the detectors controlled by two different actuators. Slight non-angular misalignment in the lens positioning and/or difference in behavior between the two actuators as well as slightly offset detector positioning affects the  $R_C$ data and this is reflected in figures b) and c) where slightly less coincidences are measured when fixed detector is at A than at B in spite of B being thinner than A by $4 \mu m$. Point to be noted is that theoretical simulations assume all slits to have equal widths and the slit B being centered on a perfectly Gaussian incident laser beam. Thus, in theory, slit B always has maximum $R_S$ as well as $R_C$. Thus difference in slit dimensions as well as small lens and detector misalignment cause some difference between the normalized theory and experimental graphs. However, the spatial correlation exhibited by our scheme is emphatic and as high as 0.9 of a possible maximum of 1.0 and is almost independent of the small inherent and unavoidable experimental non-idealness.

\section{Discussion}
We show a novel approach of generating photon pairs in SPDC which have intrinsic spatial correlation. We demonstrate a configuration which approximates the behavior of two three-dimensional quantum systems. Our approach can be used for implementation quantum information protocols which require higher dimensional quantum systems. While it has previously been shown how to encode a qutrit using a system of three slits similarly in the Ref.~\cite{Kolenderski2011}, our approach provides direct access to two correlated qutrits by modulating the laser profile of the SPDC setup.

We demonstrated a novel and simple approach to generate higher dimensional entanglement  paving a possible route towards quantum communication and information processing and fundamental entanglement  studies, using higher dimensional entangled photon states.

\section{Acknowledgments}
We acknowledge Eneet Kaur for assistance in initial calculations and especially S.N.Sahoo for technical assistance and his generous help during various stages of the project. TJ acknowleges support by the Natural Sciences and Engineering Research Council of Canada(NSERC), the Canadian Institute for Advanced Research (CIFAR), and Industry Canada.
PK acknowledges support by National Laboratory of Atomic, Molecular and Optical Physics, Torun, Poland for support, Foundation for Polish Science (FNP) (project First Team co-financed by the European Union under the European Regional Development Fund); Ministry of Science and higher Education, Poland (MNiSW) (grant no.~6576/IA/SP/2016); National Science Center, Poland (NCN) (Sonata 12 grant no.~2016/23/D/ST2/02064).\\

%\appendix

\bibliographystyle{apsrev4-1}
\bibliography{Arxiv}

%\bibliographystyle{apsrev4-1}
%\bibliography{ThreeSlits}

\section{Appendices}
%\vspace{\baselineskip}
%\textbf{\huge{Appendix}}

\subsection{Image transfer formalism}
There are certain assumptions which go into the image transfer formulation. All calculations are done in two dimensions which are the beam propagation direction "$z$" and the transverse "$x$" direction. We are assuming that the slit height can be approximated to be infinitely long compared to the slit width. Scalar field paraxial approximations for a thin lens are used. The center of the crystal is assumed to be perfectly at twice the focal length from the lens. 

The system of lenses and mirrors transfers the image of the slits to the crystal. The center of the crystal is at 2f from the thin lens. Its transfer function is given by:
\begin{multline}
H(x_i, z_i ; x_o, z_o) = \\ =\int_{-R}^{R}\frac{
e^{\frac{1}{2} \frac{i k x_o^2}{z_o}} e^{\frac{1}{2} i k \left( \frac{1}{z_o}+\frac{1}{z_i}-\frac{1}{f} \right) x_l^2} e^{- i k \left(\frac{x_o}{z_o} + \frac{x_i}{z_i} \right) x_l}
}
{\lambda^2 z_o z_i} \ dx_l.
\label{eq:transfer:function}
\end{multline}
Here ($x_0, z_0$) are the coordinates of the object whose image is to be transferred to a location ($x_i,z_i$) and $\lambda$ is the wavelength of the incident beam. Thus, in order to transfer the scalar field $U(x_0;z_0)$ to $U(x_i;z_i)$, we would use the following transformation equation and the  scalar field after the lens would then be given as:
\begin{align}
U(x_i; z_i) &= \int_{-\infty}^{\infty} dx_0 U(x_0; z_0) H(x_i, z_i ; x_0, z_0) %\\
%& =  \int_{-\infty}^{\infty} dx_o U(x_o, z_o)\int_{-R}^{R}\frac{
%e^{\frac{1}{2} \frac{i k x_o^2}{z_o}} e^{\frac{1}{2} i k \left( \frac{1}{z_o}+\frac{1}{z_i}-\frac{1}{f} \right) x_l^2} e^{- i k \left(\frac{x_o}{z_o} + \frac{x_i}{z_i} \right) x_l}
%}
%{\lambda^2 z_o z_i} \ dx_l
\label{eq:transformation}
\end{align}
It yields the image transferred at crystal position. The integration in \eqref{eq:transfer:function} is done analytically whereas the final integration in \eqref{eq:transformation} is evaluated by numerical means using \textit{Mathematica 11}. Thus instead of using uniform top hat functions as representations of the slit profiles, we have used the lens transfer formulation to transfer the image of the slits to the center of the crystal when the slits are illuminated by a Gaussian.

\subsection{Simulations to determine the optimal set of parameters for high resultant spatial correlations}
\begin{figure}[h]
	\centering
	\includegraphics[width=0.95\columnwidth]{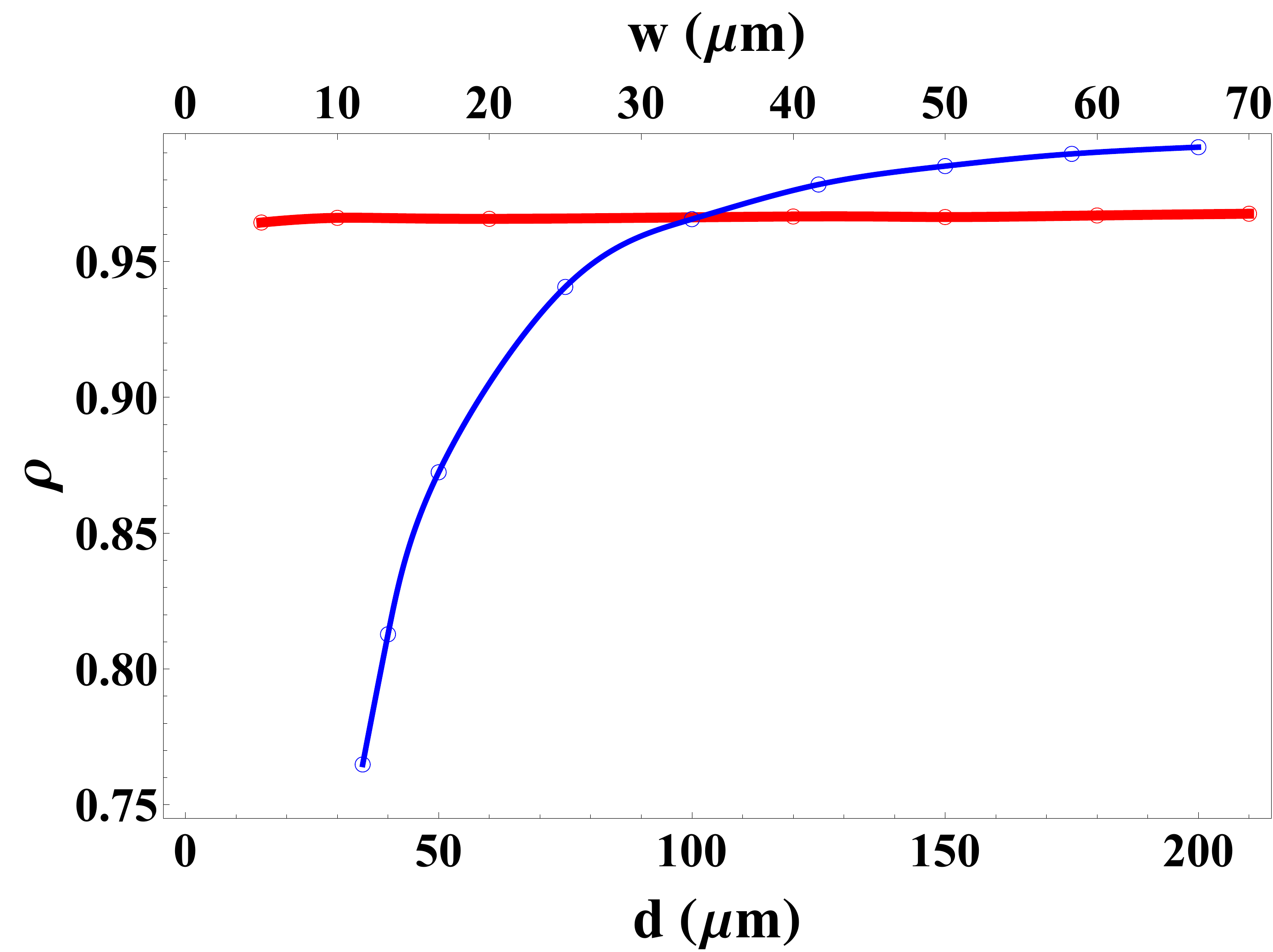}
	\caption{Variation of Pearson coefficient $\rho$ with increasing slit width and increasing inter-slit distance respectively. Simulations have been done by varying one parameter while keeping the other parameters constant. When slit width $w$ is varied, inter-slit distance is kept fixed at $100 \mu m$ whereas when inter-slit distance is varied, slit width $w$ is kept constant at $30 \mu m$. The crystal length $L_Z$ has been kept fixed at $10 mm$ for both these simulations.}
	\label{fig:pearson:slits}
\end{figure}

First, we varied the slit width keeping other parameters constant. For slit widths ranging from $5 \mu m$ to $40 \mu m$, the Pearson coefficient remained around $0.96$, see \figref{fig:pearson:slits} red curve, which indicates that for these conditions, choosing a slit width in the above range should be sufficient. We decided to choose $30 \mu m$. 
Next, we varied the inter-slit distance from $50 \mu m$ to $200 \mu m$. The Pearson coefficient is found to increase with increasing distance between the slits kept at a constant slit width as can be seen in \figref{fig:pearson:slits} blue curve. The conclusion is that when the slits are more separated, the overlap between them goes down, as a result of which the point to point correlation increases. However, a $200 \mu m$ interslit distance would entail a much bigger incident pump beam which could again lead to less throughput so we decided to choose the $100 \mu m$ interslit distance as a compromise between throughput and correlation coefficient. \\

Before choosing the crystal length, we also simulated for different crystal lengths keeping slit and beam parameters constant.  If we consider the intensity weight function associated with phase matching, we have along the longitudinal axis $sinc(\Delta k_z \times L_z/2)$. As $L_z$ increases the momentum uncertainty of the photon pairs inside the crystal also increases. As a result the correlation between the two down-converted photons decreases. This was substantiated by the simulations which showed a steady decrease in Pearson coefficient as crystal length was increased from $5 mm$ to $100 mm$. On the other hand the thinner crystals give lower pair production rate.  Thus, we selected a crystal length which is not too short but also yields an expected high correlation coefficient i.e. $10 mm$. The transverse length of the crystal needs to be larger than the extent of the transverse pump profile and was chosen to be $5 mm$.  

\begin{figure}[H]
	\centering
	\includegraphics[width=0.95\columnwidth]{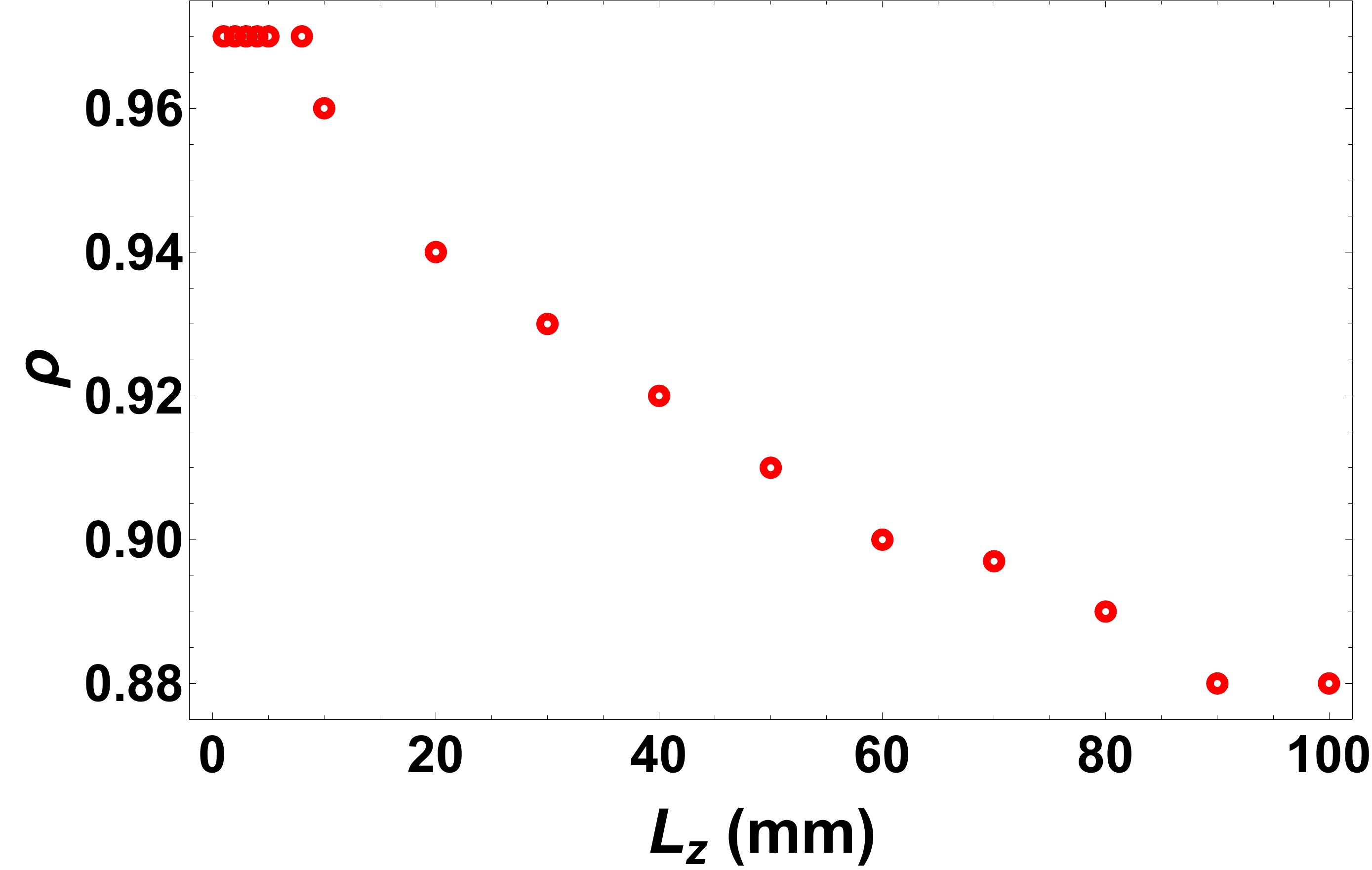}
	\caption{Variation of $\rho$ with $L_z$. As crystal length increases, the Pearson coefficient is seen to decrease.}
	\label{fig:pearson:length}
\end{figure}

\subsection{Comparison between experimental and theoretical triple slit image transfer}

\begin{figure}[h]
	\includegraphics[width=\columnwidth]{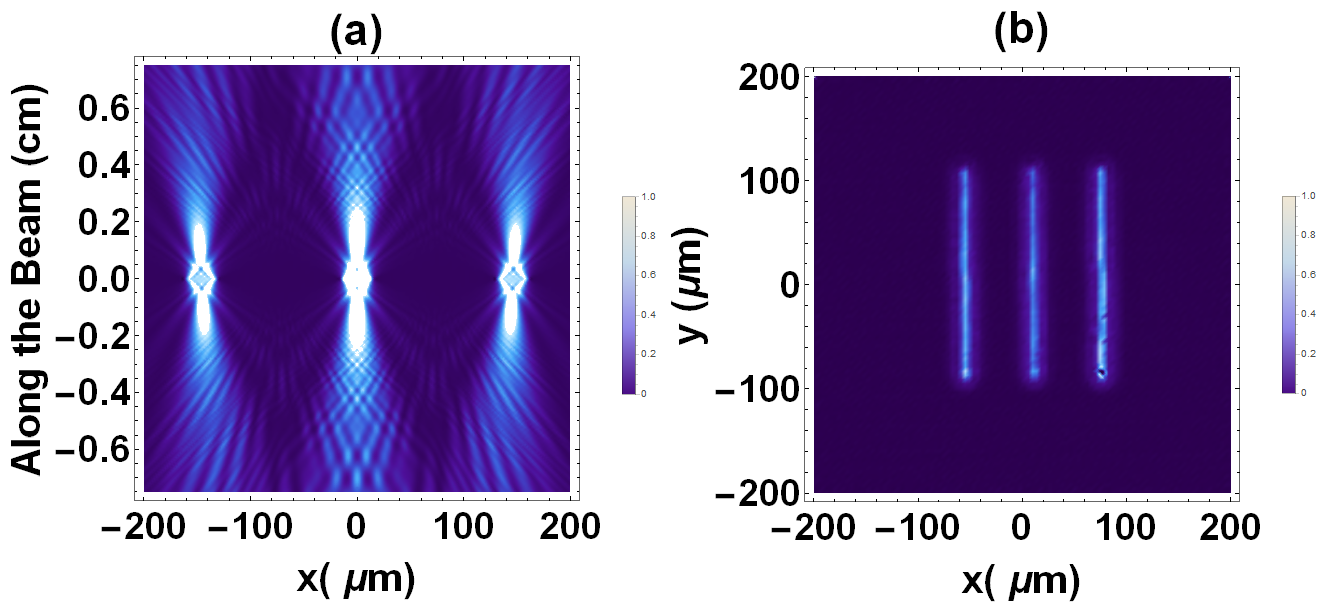}
	\caption{Figure on left shows the theoretically simulated pump profile. When the lens is used for image transfer experimentally, a magnification is introduced in the system, which has also been incorporated in theory. While the y-axis denotes the crystal length along beam propagation direction, the x-axis denotes the image along the transverse crystal direction. The figure on the right is the experimentally measured image of the modulated pump at the position corresponding to center of the crystal using a lens.}
	\label{fig:image transfer}
\end{figure}

Figure \ref{fig:image transfer} shows an example of the image transferred to the center of the crystal comparing experimentally obtained images with theoretically simulated ones with a triple slit modulated pump profile. Figure on left shows the image as a function of position in the crystal along beam propagation direction.

\end{document}